# Quantifying the Multivariate ENSO Index (MEI) coupling to $CO_2$ concentration and to the length of day variations

## A. Mazzarella[1], A. Giuliacci [1,2] and N. Scafetta [3]

*(1) Meteorological Observatory*
*Department of Earth Science – University of Naples Federico II*
*Largo S. Marcellino 10, 80138 Naples, Italy*

*(2) Epson Meteo Center, via de Vizzi 93/95,*
*20092 Cinisello Balsamo, Milan, Italy*

*(3) Active Cavity Radiometer Irradiance Monitor (ACRIM) Lab,*
*Coronado, CA 92118, USA*
*Duke University, Durham, NC 27708, USA*

## Abstract

The El Niño Southern Oscillation (ENSO) is the Earth's strongest climate fluctuation on inter-annual time-scales and has global impacts although originating in the tropical Pacific. Many point indices have been developed to describe ENSO but the Multivariate ENSO Index (MEI) is considered the most representative since it links six different meteorological parameters measured over the tropical Pacific. Extreme values of MEI are correlated to the extreme values of atmospheric $CO_2$ concentration rate variations and negatively correlated to equivalent scale extreme values of the length of day (LOD) rate variation. We evaluate a first order conversion function between MEI and the other two indexes using their annual rate of variation. The quantification of the strength of the coupling herein evaluated provides a quantitative measure to test the accuracy of theoretical model predictions. Our results further confirm the idea that the major local and global Earth-atmosphere system mechanisms are significantly coupled and synchronized to each other at multiple scales.

**INTRODUCTION**

El Niño-La Niña is the strongest quasi-oscillatory pattern observed in the climate system and it is coupled to numerous climatic systems. Numerous empirical and theoretical studies have attempted to discover its multivariate influences and to model it in general circulation models (see for example: Graf and Zanchettin, 2012 and the literature there referenced).

However, current general circulation models (GCMs) do not reproduce well the patterns observed in climatic data such as trends and cycles at multiple time scales (Douglass et al. 2007; Scafetta 2010, 2012b; Spencer and Braswell, 2011). The models also fail to forecast the summer from the preceding winter and vice-versa, and are unable to accurately simulate and predict some important circulation phenomena such as the quasi biennial oscillations (QBO) and the El Niño/La Niña-Southern Oscillation (ENSO). These major climate variations are supposed to be generated by a not-well understood internal dynamics (Meehl et al., 2011), although a contribution from astronomical harmonic forcings cannot be excluded (Wang et al., 2012). In general, numerous uncertainties affect our understanding of climate dynamics (Curry and Webster, 2011).

The climate system is made of a set of subsystems coupled to each other and behaves as a complex network of coupled non-linear oscillators, which synchronize to each other (Tsonis et al. 2008; Wyatt et al. 2011). For example, Scafetta (2010, 2012a) has shown that all major global, hemispheric, land and ocean surface temperature records are characterized by no less than 11 common frequencies from a period of 5 years to 100 years that match equivalent astronomical cycles of the heliosphere and terrestrial magnetosphere. Mazzarella and Scafetta (2012) have shown that the North Atlantic Oscillation index (NAO), the global ocean temperature, length of day, and a record of historically observed mid-latitude auroras present common quasi 60-year oscillations since 1700, which also suggests an astronomical origin of major climatic oscillations. Many other examples stress the importance of studying and directly quantify the strength of the coupling among alternative physical observables. Indeed, empirical modeling of climate change might have higher predicting power than traditional analitical models. Once that the strength of the complings among the climate subsystems is properly quatified, it may be possible to evaluate how well proposed physical models can reproduce them. This testing process would eventually yield better and more accurate theoretical models.

Bacastow (1976) found that atmospheric carbon dioxide record is correlated to the Southern Oscillation Index (SOI), which indicates that a component of the change in the rate of CO2 removal is regulated by the southern tropical wind and ocean oscillations. More recently, Zheng et al. (2003) concluded that ENSO events, the changes in the length of day (LOD), and the global atmospheric angular momentum are correlated. However, Bacastow and Zheng et al. simply evaluated the correlation coefficients between two records and their reciprocal time lag. Indeed, a more quantitative relation would be more useful because it can be more directly used to test the accuracy of the models.

In the following we study the mutual relations among: the recent availability of monthly data of Multivariate Enso Index (MEI) that measures the El Niño Southern Oscillation (ENSO), which is the Earth's strongest natural climate fluctuation on inter-annual time-scales; the atmospheric CO2 concentration measured at Mauna Loa; and the length of day (LOD), which is a global astronomical observable. We notice that MEI is a more comprehensive index than SOI and ENSO. We use a similar mathematical methodology to study the mutual correlation and to quantify it. Finally, we discuss possible underlying geophysical phenomena that could explain the findings.

**DATA COLLECTION**

We have analyzed the monthly values of:

1) Multivariate Enso Index (MEI) (interval: 1950-2011) as computed by Wolter and Timlin (1993, 1998) (http://www.cdc.noaa.gov/people/klaus.wolter/MEI/table.html).

The record is depicted in Figure 1A. Each monthly value is centered between the preceding and subsequent month: for example, the January value represents the value centered between the December-January months and so on. MEI is a multivariate measure of the ENSO signal. It is the first principal component of six main observed variables over the tropical Pacific: sea level pressure, zonal and meridional components of the surface wind, sea surface temperature, surface air temperature and cloudiness of the sky. The MEI monthly values are standardized with respect to a 1950-1993 reference period and are expressed as fractions of standard deviation for which it has a total mean equal to zero and a standard deviation equal to 1.

2) $CO_2$ (ppm) monthly concentration data measured at Mauna Loa (lat: 19°32'10" N; long: 155°34'34" W; height: 3397 m; interval:1958-2011), ( ftp://ftp.cmdl.noaa.gov/ccg/CO2/trends/CO2_mm_mlo.txt). The record is depicted in Figure 1B.

3) Length of Day (LOD) (ms) (interval: 1962-2010), i.e., the difference between the astronomical length of day and the standard length (interval: 1962–2010) (Stephenson and Morrison 1995) (ftp://hpiers.obspm.fr/eop-pc/eop/eopc05/eopc05_dailyftp://hpiers.obspm.fr/eop-pc/eop/eopc05/eopc05_daily). The record is depicted in Figure 1C.

## METHODOLOGY AND RESULTS

### a.Long-term analysis

The three curves depicted in Figure 1 appear quite different from each other. The MEI index fluctuates in an irregular way around a zero average. The $CO_2$ concentration record presents a clear upward trend due to the addition of anthropogenic gases plus a smaller annual oscillation due to the physical asymmetry between the Northern and Southern hemispheres. The LOD decreases and presents a clear annual cycle plus an apparently cyclical modulation with period of about 18-20 years, which, perhaps, may be astronomically induced by the 18.6-year solar-lunar nodal cycle or other astronomical cycles (Douglass et al. 2007) (we do not discuss this issue further in this paper). The dynamical patterns observed in the original records depicted in the figure would suggest that the three records are strongly uncorrelated: the cross-correlation between: MEI-$CO_2$ gives r = 0.084; MEI-LOD gives r = 0.011; $CO_2$-LOD gives r = -0.69, which is negative and large only because one record ($CO_2$) has an upward trend and the other (LOD) has a downward trend during the given period. The linear regression analysis parameters for the three records are summarized in Table 1.

The above results would suggest that no simple relation exists among the three records.

### b. Seasonal annual cycle analysis

MEI does not present any evident regular annual cyclical variability. Each of the six time series is "normalized by [first] computing the bimonthly anomalies from the respective 135-year averages" (Wolter and Timlin, 2011). On the contrary both $CO_2$ and LOD present a clear annual cyclicity. LOD follows quite closely the annual cycle with maximum values during the winter and minimum values during the summer, while $CO_2$ annual cycle presents maximum values in the spring (April-June) and minimum values during September and October. See Figure 2 where for visual convenience we compare the two records from 1978 to 1985.

The $CO_2$ annual cyclicity can be easily explained observing that from October to April the Northern Hemisphere (NH) cools while the Southern Hemisphere (SH) warms. Because most land is located in the NH while most ocean is located in the SH, the $CO_2$ atmospheric concentration is expected to increase from October to April because both the plants in the NH and the warmer ocean in the SH would uptake less $CO_2$ from the atmosphere. The opposite situation occurs from May to September.

The annual seasonal variation in LOD has been first extensively discussed by Lambeck (Lambeck 1980), who identified the variable zonal wind circulation as the cause of LOD seasonal cycle. The annual LOD cyclicity is related to the annual temperature cycle. In fact, during the Northern hemisphere winter the Earth is at its closest distance from the Sun and the incoming total solar irradiance is on average about 40 W/m$^2$ larger than during the summer. So, perhaps, during the winter the over all temperature of the entire planet (ocean plus atmosphere) increases causing a change in wind velocities that may result in exchanges of angular momentum between the atmosphere and the Earth (Rosen and Salstein 1985). LOD also presents a clear 6 month cycle that is a sub-harmonic of the annual cycle that may be related to a solar tidal harmonics whose detailed analysis is left to another study.

## c. Annual rate variation analysis

The lack of a linear correlation among the three variables should not be taken to mean that these variables are not coupled. Indeed, a strong coupling may exist, but it is simply nonlinear. Herein, we investigate whether a better correlation exists among MEI, LOD and $CO_2$ using their annual rate variation function.

We proceed in the following way. First we process the signals by eliminating the large seasonal variations identified in $CO_2$ and in LOD by annual differentiating the signals. That is, we compute the difference between the value of January 1963 and that of January 1962, between the value of February 1963 and that of February 1962 and so on, for each month and for each year. The value is centered in the average of the chosen interval: for example, the difference between Jan/1963 and Jan/1962 will be centered in 0.5*(1963.04+1962.04) = 1962.54. At the end, we obtain a monthly series of $CO_2$ and LOD annual rate variation. Figure 3A depicts the time plot of MEI and of $CO_2$ annual rate variation record. Figure 3B depicts the time plot of MEI and of LOD annual rate variation record. The processed MEI index is rescaled by using the linear conversion relation depicted in Figure 4.

We observe a very good correlation among the extreme values, such as during the strong El Niño event of 1998. Note the good synchrony occurring during the El Niño events such as in 1965, 1972, 1983, 1987 and 1998 and with corresponding La Niña events.

In the figure, $CO_2$ lags MEI by 3 months (best correlation coefficient: r = 0.49) while LOD lags MEI by 4 months (best correlation coefficient: r = -0.34). Both correlation coefficients are highly significant (P(|r|>|r$_0$|)<0.01). This result would suggest that LOD and $CO_2$ changes are driven by MEI oscillations.

It appears that 3 months after all El Niño events, the $CO_2$ rate reaches a peak with the exception of the interval around 1991, and all La Niña events are followed by lowest values of $CO_2$. A similar argument can be repeated for the LOD annual rate index.

To explain the absence of synchrony around 1991, it is worth noting that a violent eruption of Mount Pinatubo (15°08'30"N; 120°21'00" E; 1745 m asl), located in the same tropical latitude and upwind of Mauna Loa, began in June 1991. It was the second largest eruption of the 20$^{th}$ century and has been classified with a (Volcanic Explosivity index) VEI = 6. For a few years after a major volcanic eruption (i.e. when there is an abundance of sulphate aerosols in the atmosphere) heterotrophic respiration decreases due to a lowering of the Earth's surface temperature and the productivity of forest ecosystems increases under enhanced diffuse radiation. Both processes lead to a negative anomaly in $CO_2$ growth rate that may explain the absence of synchrony between MEI and $CO_2$ in the 1991-1993 interval (Patra et al. 2005).

Figures 4A and 4B show scatter-graphs of the MEI index against the $CO_2$ and LOD annual rate variations, respectively. Figure 4A suggests that in first order approximation the $CO_2$ annual rate variation can be obtained from MEI by approximately multiplying the latter by K = 0.55 ppm/year. Figure 4B suggests that in first approximation the LOD rate variation can be obtained from MEI by approximately multiplying the latter by K = -0.27 ms/year. The two first order conversion functions are depicted in Figures 4a and 4B and are used to prepare the graphical comparisons depicted in Figures 3A and 3B.

**DISCUSSION AND CONCLUSIONS**

ENSO is the Earth's strongest natural climate fluctuations on inter-annual time-scales. It is a complex atmospheric and oceanographic phenomenon that has profound economic and social consequences (Wang and Fiedler, 2006). However, ENSO is best described by MEI that combines six representative meteorological variables measured in the tropical Pacific.

Current GCMs are not able to reproduce or forecast ocean oscillations such as ENSO events. This failure may be due to poor understanding of the ocean oscillations, their physical mechanisms and true forcings (Scafetta 2010, 2012c; McLean 2009). It has been proposed that ENSO is, at least to some degree, a stable mode or oscillation triggered by random disturbances (Philander and Fedorov, 2003). ENSO oscillations may also be interpreted in terms of a self-organized critical state (Mazzarella and Giuliacci 2009). However, complex astronomical and tidal cyclical forcings, today ignored in the climate models, may be significantly involved in the process (Scafetta, 2010, 2012a, 2012b; Wang et al., 2012). Thus, it is important to analyze the data in details to identify all physical mechanisms that may be involved in the process.

Herein, we have studied three geophysical indexes: MEI, LOD and CO2 records. We have shown that the LOD and CO2 annual rates are well correlated to MEI. In Figure 4 we have quantified the conversion factors and showed the good agreement in Figure 3.

The highest values of MEI show a direct and an inverse relationship with the highest values of $CO_2$ and LOD annual rate occurring after just a few months, respectively (Figure 3). Since the highest values of MEI represent El Niño events, the results here obtained show the influence of El Niño on $CO_2$ and on LOD. It is worthy noting that El Niño events occur in correspondence with an increase of sea surface temperature and a weakening of easterly trade winds (Wang and Fiedler 2006; Deser and Wallace 1990; Wallace 1998). But, a weakening of easterly winds causes an increase of zonal wind (Mazzarella, 2008, 2009) that, like a torque, causes an acceleration of the Earth's rotation, i.e., a decrease of LOD. Equally, an increase of sea surface temperature causes a smaller solubility of $CO_2$ in the ocean and so a higher concentration in the atmosphere.

We propose that the nature and the magnitude of these correlations herein evaluated should be used to validate any analytical model attempting to reproduce the climate system in its effects and components.

|        | m               | n             |
|--------|-----------------|---------------|
| MEI    | 0.012 ± 0.003   | -0.18 ± 0.09  |
| $CO_2$ (ppm) | 1.48 ± 0.01 | 309.7 ± 0.3  |
| LOD (ms) | -0.042 ± 0.002 | 2.93 ± 0.06 |

**Table 1.** Values of linear regression calculated using the function Y(t) = m (t-1962) + n from 1962 to 2010 of MEI with $CO_2$ and LOD annual rate indexes.

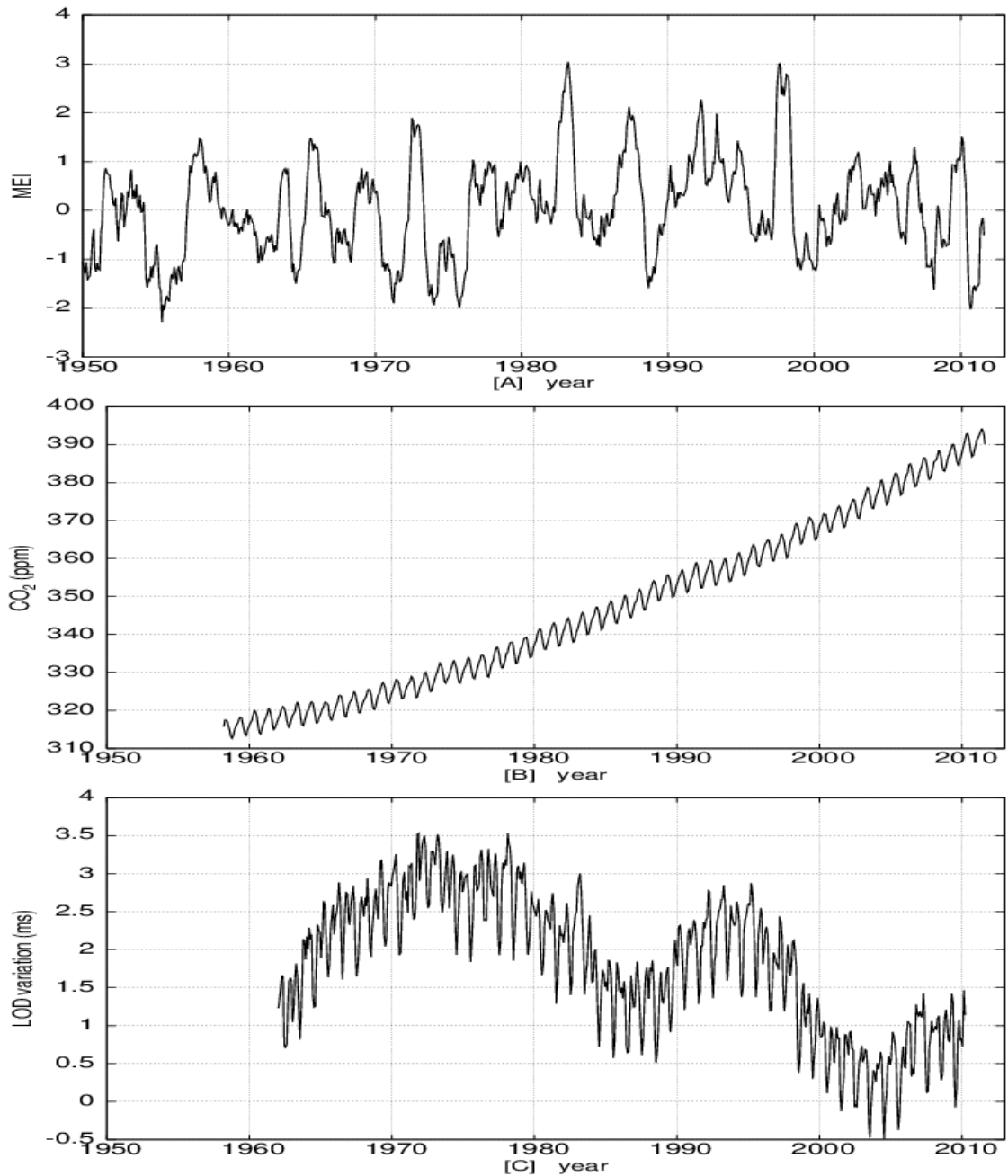

**Figure 1.** [A] Time plot of monthly MEI values (interval: 1950-2011). [B] Time plot of monthly values of $CO_2$ measured at Mauna Loa (lat: 19°32'10" N; long: 155°34'34" W; height:3397 m; interval:1958-2011). [C] Time plot of monthly LOD values (interval 1962-2010).

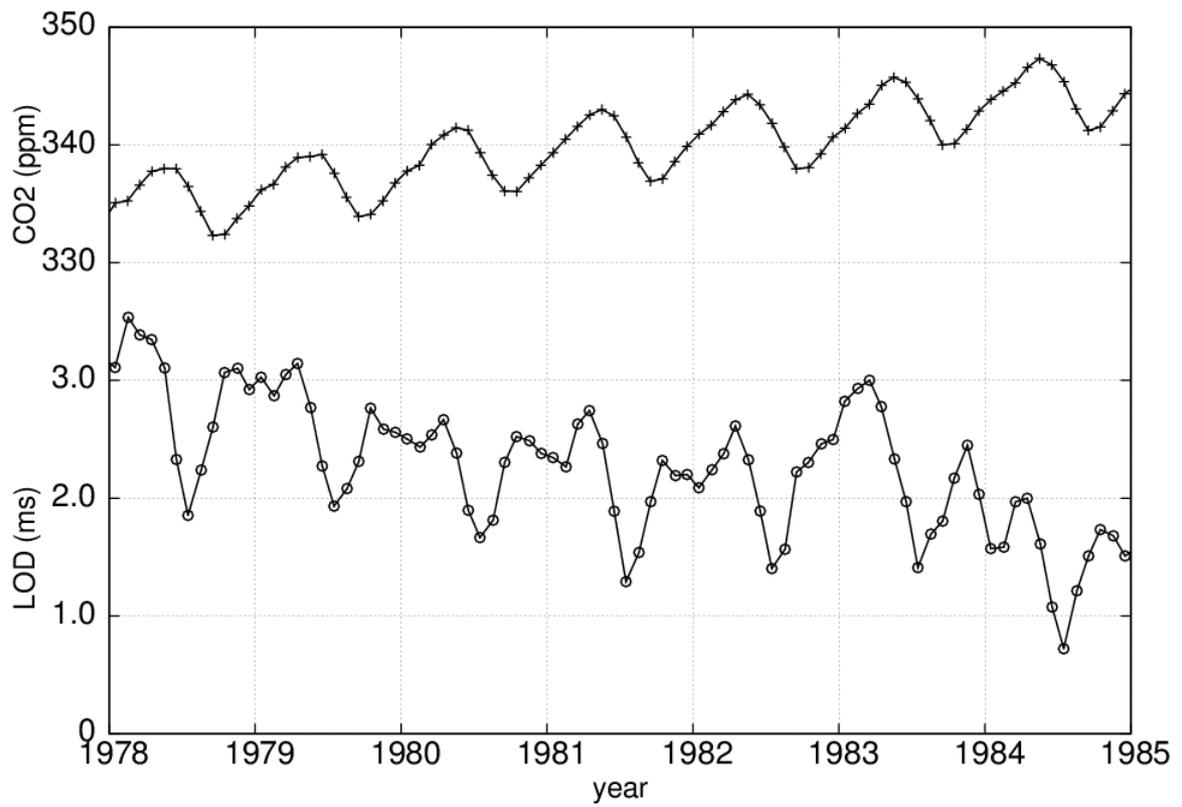

**Figure 2**. Illustration of the annual cycle of LOD and $CO_2$ from 1978 to 1985**,** other time intervals are qualitatively equivalent. The CO2 cycle lags the LOD annual cycle by about a season.

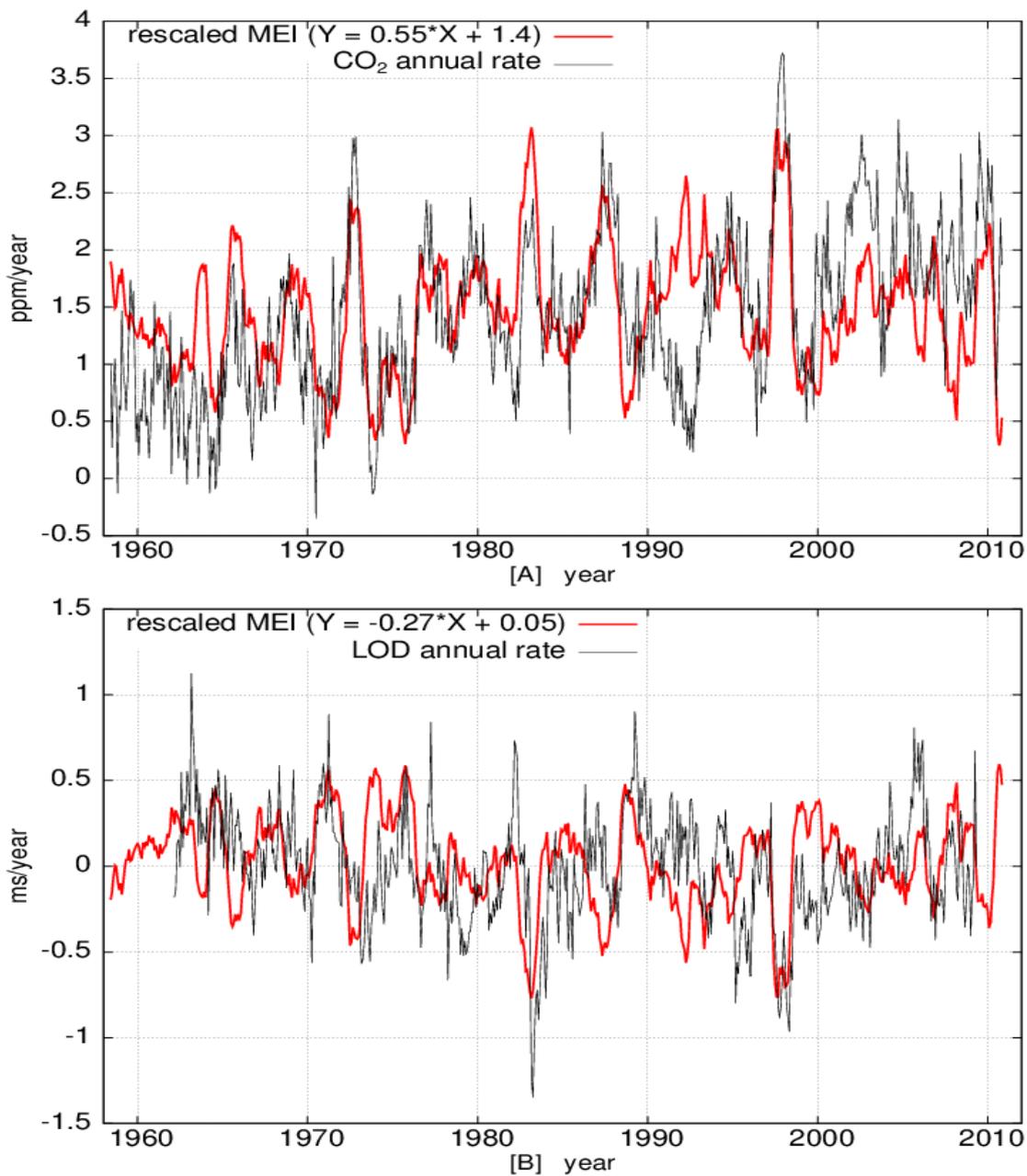

**Figure 3.** [A] Time plot of rescaled MEI index (red line) and of $CO_2$ annual rate variation. The $CO_2$ curve is shifted 3 months back for best correlation. [B] Time plot of MEI (red line) and of LOD annual rate, which is shifted 3 months back for best correlation. Note that in [B] the MEI record is not only rescaled but also flipped up-down for visually helping a reader to notice its good correlation with LOD annual rate. The CO2 and LOD curves are plotted against a rescaled MEI index according to the scatter-graph results depicted in Figure 4. Note the good correlation between the depicted curves where the larger minimum and maximum extremes usually correspond.

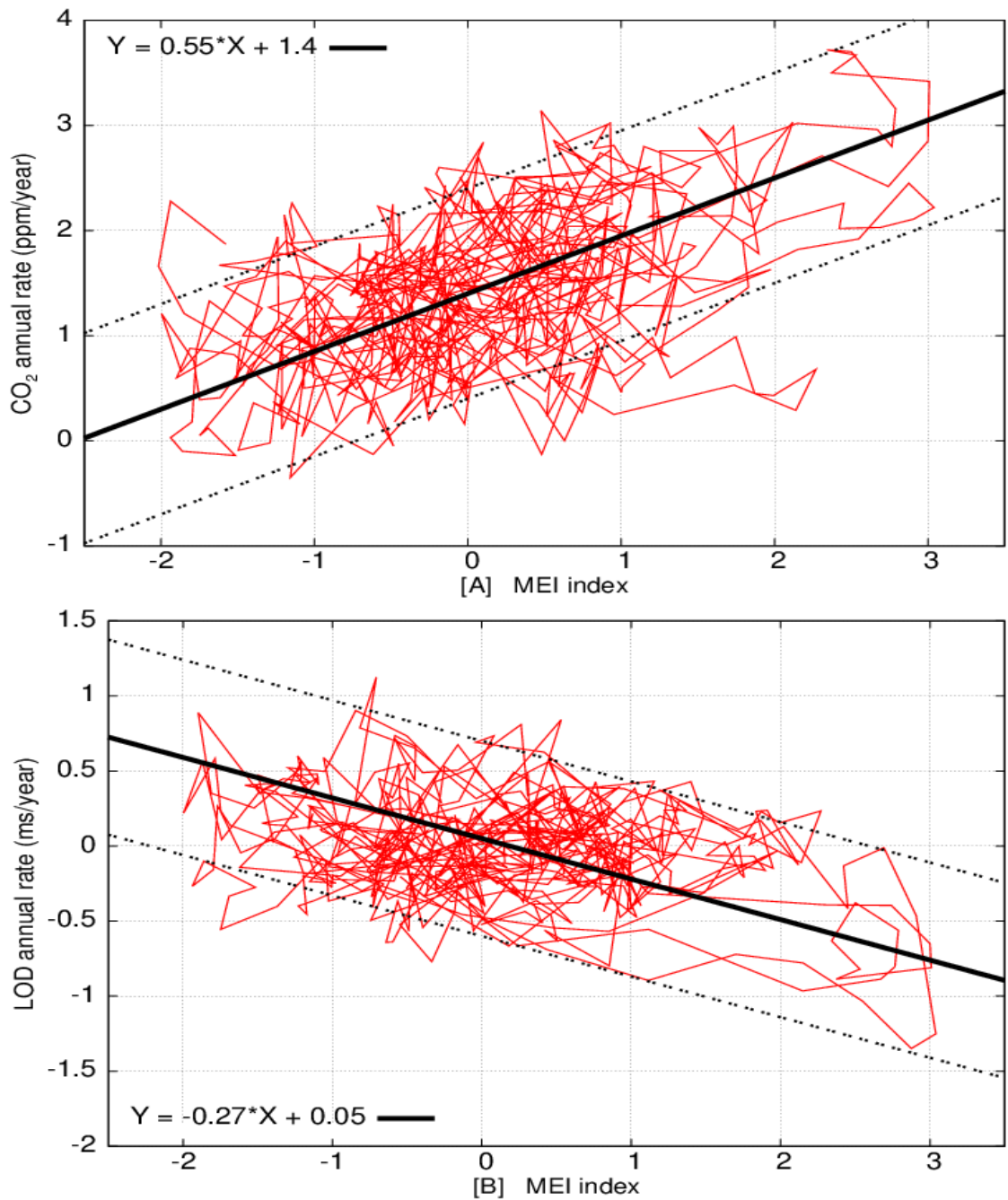

**Figure 4.** [A] Scatter-graphs of the MEI index against the annual rate of $CO_2$ (the $CO_2$ rate index is shifted back by 3 months). [B] Scatter-graphs of the MEI index against the annual rate of LOD (the LOD rate index is shifted back by 4 months). The figures report also the proposed optimal first order conversion functions that have been used in Figures 3A and 3B, respectively.